\documentclass[twocolumn,amsmath,amssymb]{revtex4}
\pdfoutput=1
\usepackage{latexsym}
\usepackage{amsmath}
\usepackage{soul,xcolor}
\usepackage{times}
\usepackage{amssymb}
\usepackage{fancyheadings}
\usepackage[T1]{fontenc}
\usepackage[utf8]{inputenc}
\usepackage{url}
\usepackage{hyperref}
\usepackage{color}
\usepackage{graphicx}
 \usepackage{epsfig}
\usepackage{mathptmx}
\usepackage{bm}
\usepackage{array}
\usepackage{algorithm}
\usepackage{algorithmic}
\usepackage{url}
\usepackage{hyperref}
\usepackage{color}
\usepackage{soul,xcolor}
\setstcolor{red}

\usepackage{lscape}
\usepackage{draftcopy}

\begin{document}
\title{Hysteresis and synchronization processes of Kuramoto oscillators on high-dimensional
  simplicial complexes with   the competing simplex-encoded couplings}
\author{Malayaja Chutani$^1$, Bosiljka Tadi\'c$^{2,3}$, and Neelima Gupte$^{1,4}$}
\affiliation{$^1$Department of Physics, Indian Institute of Technology Madras, Chennai 600036, India}
\affiliation{$^2$Department of Theoretical Physics,  Jo\v zef Stefan Institute,
Jamova 39, Ljubljana, Slovenia}
\affiliation{$^3$Complexity Science Hub Vienna - Josephstadter Strasse 39, Vienna, Austria}
\affiliation{$^4$Complex Systems and
  Dynamics Group, Indian Institute of Technology Madras, Chennai 600036, India}

\vspace*{3mm}
\date{\today}

\begin{abstract}
\noindent Recent studies of dynamic properties in complex systems point out the profound impact of hidden geometry features known as simplicial complexes, which enable geometrically conditioned many-body interactions. Studies of collective behaviours on the
controlled-structure complexes can reveal the subtle interplay
of geometry and dynamics. Here, we investigate the phase
synchronisation (Kuramoto) dynamics
under the competing interactions embedded on 1-simplex (edges) and
2-simplex (triangles) faces of a homogeneous 4-dimensional simplicial
complex. Its underlying network is a 1-hyperbolic graph with the assortative correlations among the node's degrees and the spectral dimension that
exceeds $d_s=4$. By solving numerically the set of coupled equations
for the phase oscillators associated with the network nodes, we determine the time-averaged system's order parameter to characterise the synchronisation level. 
In the absence of higher interactions, the
complete synchrony is continuously reached with
the increasing positive pairwise interactions ($K_1>0$), and a partial
synchronisation for  the negative couplings ($K_1<0$) with no apparent
hysteresis.  Similar behaviour occurs in the degree-preserved
randomised network. In contrast, the synchronisation is absent for
the negative pairwise coupling  in the entirely random graph and 
simple scale-free networks. 
Increasing the strength $K_2\neq 0$ of the triangle-based interactions
gradually 
hinders the synchronisation promoted by pairwise couplings, and the non-symmetric hysteresis loop opens with an abrupt desynchronisation transition towards the
$K_1<0$ branch.  However, for substantial triangle-based interactions,
the frustration effects prevail, preventing the complete synchronisation, and the abrupt transition disappears. These findings shed new light on the
mechanisms by which the high-dimensional simplicial complexes in
natural systems, such as human connectomes, can modulate their native synchronisation processes.
\end{abstract}
\maketitle
\section{Introduction\label{sec:intro}}
In complex systems, collective dynamics  is a marked signature of emerging
properties related to complex structure, studied by mapping onto networks
\cite{CN_explPhenAArev2019}. The synchronisation of oscillator systems
is a paradigmatic stochastic process for study the emergence of coherent behaviour in many natural and laboratory
systems \cite{Rodrigues2016_sync_review}.
In recent years, research
focuses on the system's hidden geometry features
\cite{NetGeometry2021,NetGeom2018SD} and their impact on dynamics. Notably, new
dynamical phenomena appear that can be related to the higher-order connectivity
and interactions supported by the system's hidden geometry, which is
mathematically described by simplicial complexes \cite{SC_Arenas_NatureComm,HOC-PhysRep2020,SC_we_Entropy2020,SC_we_EPL2020,Sync_Arenas_PRL,DdimSC_boccaletti2020,HOC_contagAA_PRX2020}.

A formal theory of the simplicial complexes of graphs 
\cite{jj-book,ZhaoMaleticSimpCompBook,BeaumontJohnR1982AitQ} defines a simplicial complex as a structure consisting
of different simplexes, e.g., $n$-cliques, which share one or more
common nodes representing a geometrical face of the implicated
simplexes.  For example, an  $n$-clique is a full graph of n nodes, and
its  faces are simplexes of the
order $q=0,1,2,3,... q_{max}$, where $q_{max} =n-1$ defines the dimension of
the simplex. Hence, the dimension of the simplicial complex is defined by order of the largest simplex that it contains. 
Recent studies revealed simplicial architecture in networks mapping many complex systems
from the brain\cite{we_Brain2016,we-Brain_SciRep2019,we-Brain_SciRep2020,Brain_balanceSciRep21}, designed materials \cite{AT_Materials_Jap2016}
 and physics problems \cite{we_PRE2015,we_Qnets2016,Geometries_QuantumPRE2015,Chutani2020TSNets}, to structures emerging from online social endeavours \cite{we_Tags2016} and large-scale social networks \cite{HOC_evolution2020,HB-knowNets-BT2019}.
Such simplicial structure naturally underlies many-body interactions
\cite{HOC-PhysRep2020}. 
 However, revealing the mechanisms by which such high-dimensional  simplicial complexes
 determine collective dynamics in these complex systems
 represents a challenging problem.

In this context, the generative models of high-dimensional simplicial
complexes of a controlled structure are of great importance \cite{SC_we_SciRep2018,SC_we_PRE2020,we-ClNets-applet,SC_Sfboccaletti2021}. 
For example, in the model of self-assembly of cliques of different
size developed in Ref.\ \cite{SC_we_SciRep2018}, the assembly is
controlled by two factors. These are the geometric compatibility of
the attaching clique's faces with the once already built-in the
growing structure, and the chemical affinity towards the addition of new
nodes.   Varying the chemical affinity, one can grow different
structures from sparsely connected cliques that share a single node to
the very dense structure of large clique sharing their most prominent
sub-clique, see the online demo on \cite{we-ClNets-applet}. Moreover, the architecture of simplicial complexes manifests on several unique properties of the underlying network
(1-skeleton of the simplicial complex), that can affect the pairwise interaction, see
section\ \ref{sec:nets} for details. For example, the network's
spectral dimension can vary from the values
close to the tree graphs in sparsely connected cliques of any size, to
the  values $d_s\geq 4$, in the case of large densely  connected
cliques,  as shown in Ref.\ \cite{SC_we-PREspectra2019}.  %
Hence, the structure--dynamics interplay can
be expected both because of the pairwise and higher-order interactions
due to the actual architecture of simplicial complexes. 
More precisely, it has been demonstrated by studies of spin kinetics
\cite{SC_we_Entropy2020,SC_we_EPL2020}, contagious dynamics \cite{HOC_contagAA_PRX2020}, and synchronisation processes
\cite{Sync_Arenas_PRL,HOC-Synch_Ginestra2020,DdimSC_boccaletti2020} on various
simplicial complexes.  Notably, in the field-driven magnetisation
reversal on simplicial complexes
\cite{SC_we_Entropy2020,SC_we_EPL2020}, the antiferromagnetic
interactions via links of the triangle faces provide  strong geometric
frustration effects that determine the shape of the hysteresis loop. The higher-order
interactions then affect its symmetry, meanwhile the width of the
hysteresis remains strictly determined by the dimension of the
simplicial complex. In the contagious and synchronisation processes,
on the other hand, the appearance of the hysteresis loop is strictly related to the higher-order interactions.
The synchronisation has been studied extensively on a variety of networks
\cite{Arenas2008_sync_review,Rodrigues2016_sync_review}  using an
ensemble of 2-dimensional phase oscillators (Kuramoto model) with
interactions via network edges. It has been understood that the onset
of synchronous behaviour can be affected by the local connectivity and
correlations among the nodes \cite{Synch_assortativity2013}, and global features captured by the network's spectral dimension \cite{millan2019synchronization}. The nature of synchronisation transition can depend on the
process' sensitivity to the sign of interactions, time delay, and the frustration effects causing new phenomena \cite{wu2018dynamics,dai2018interplay,ha2018emergence,SYNCH_Frust_Chaos2015,SYNCH_frustr_Frontiers2016,SYNC_FrustrExplosive2018}.
 Furthermore, the presence of higher-order interactions are shown to
 induce an abrupt desynchronisation, depending on the dimension of the
 dynamical variable, and the range of couplings
 \cite{Sync_Arenas_PRL,DdimSC_boccaletti2020}. 
 It remains unexplored how the coincidental interactions of a different order,  encoded by the faces of a large simplicial complex, cooperate during the synchronisation processes.

Here, we tackle this problem by  numerical investigations of
synchronisation and desynchronisation processes among Kuramoto phase oscillators considering the leading
 interactions based on 1-simplices (edges)  and 2-simplices (triangles)
as the faces of homogeneous  4-dimensional simplicial complexes.  
The structure is grown by self-assembly of 5-cliques that preferably
share the most extensive face. The underlying graph of this simplicial complex possesses several unique features. These are hyperbolicity, assortative mixing, and a high spectral dimension, allowing complete synchronisation when the positive pairwise interaction is increased. Our results suggest that these geometrical properties, even in the absence of the higher-order interactions, can lead to new states with partial
synchronisation, mainly when the pairwise coupling is
negative, which can be attributed to frustration. 
Furthermore, these simplex-based interactions have
competing effects, leading to different patterns of synchronisation and desynchronisation. 
Remarkably, the triangle-based interactions tend to hinder the
synchronisation processes promoted by the increasing pairwise
coupling, leading to the hysteresis loop and the abrupt
desynchronisation when the fully synchronised state can be reached, i.e.,
for a moderate strength of interaction. Both the complete synchrony
and the abrupt desynchronisation disappear for strong triangle-based interactions, suggesting the dominance of geometric frustration.

In section \ref{sec:nets}, we present the relevant details of the
structure of the simplicial complex and the underlying
network. Section\ \ref{sec:dyn} introduces the dynamical model with
the simplex-based interactions and discusses the case with pairwise
interaction alone. In section\ \ref{sec:HL}, the effects of
the triangle-based interactions on the order-parameter and
hysteresis loop are shown. Section \ref{sec:concl} presents a summary and
discussion of the results.

\begin{figure*}[!]
\begin{tabular}{cc} 
\resizebox{32pc}{!}{\includegraphics{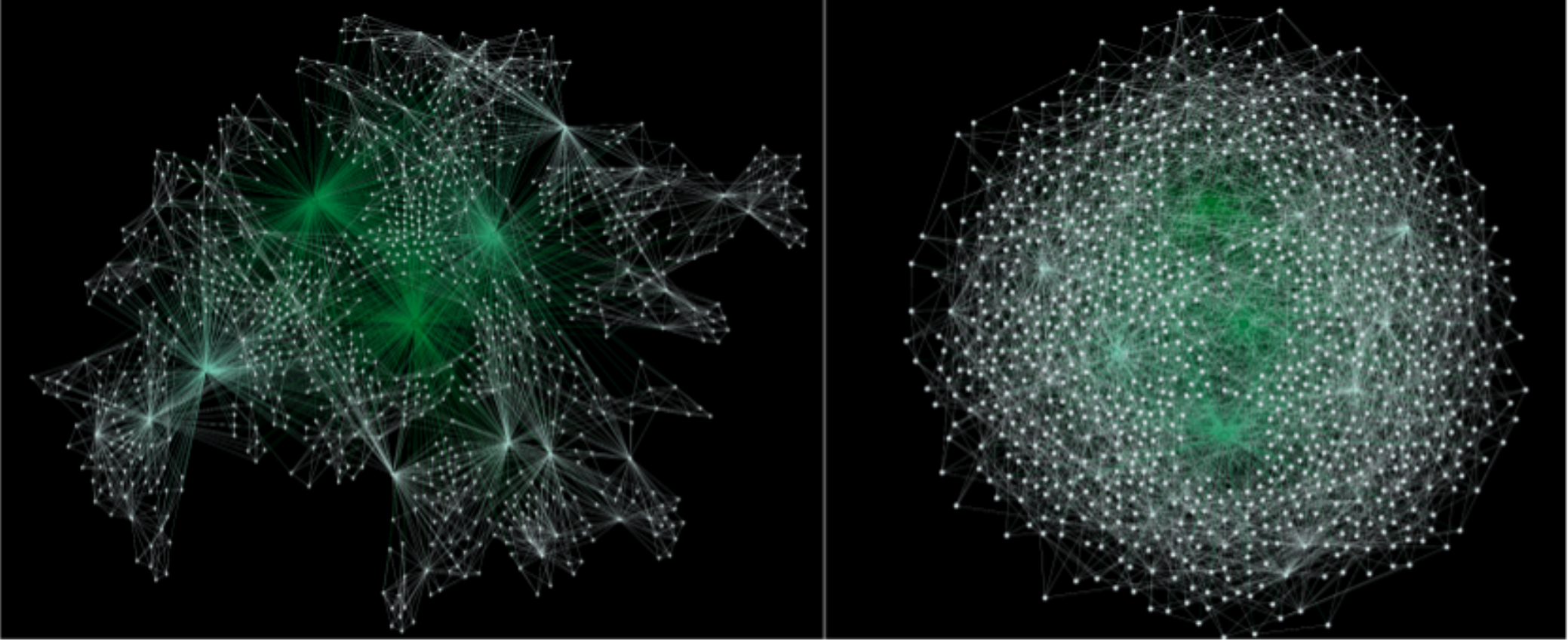}}\\
\resizebox{34pc}{!}{\includegraphics{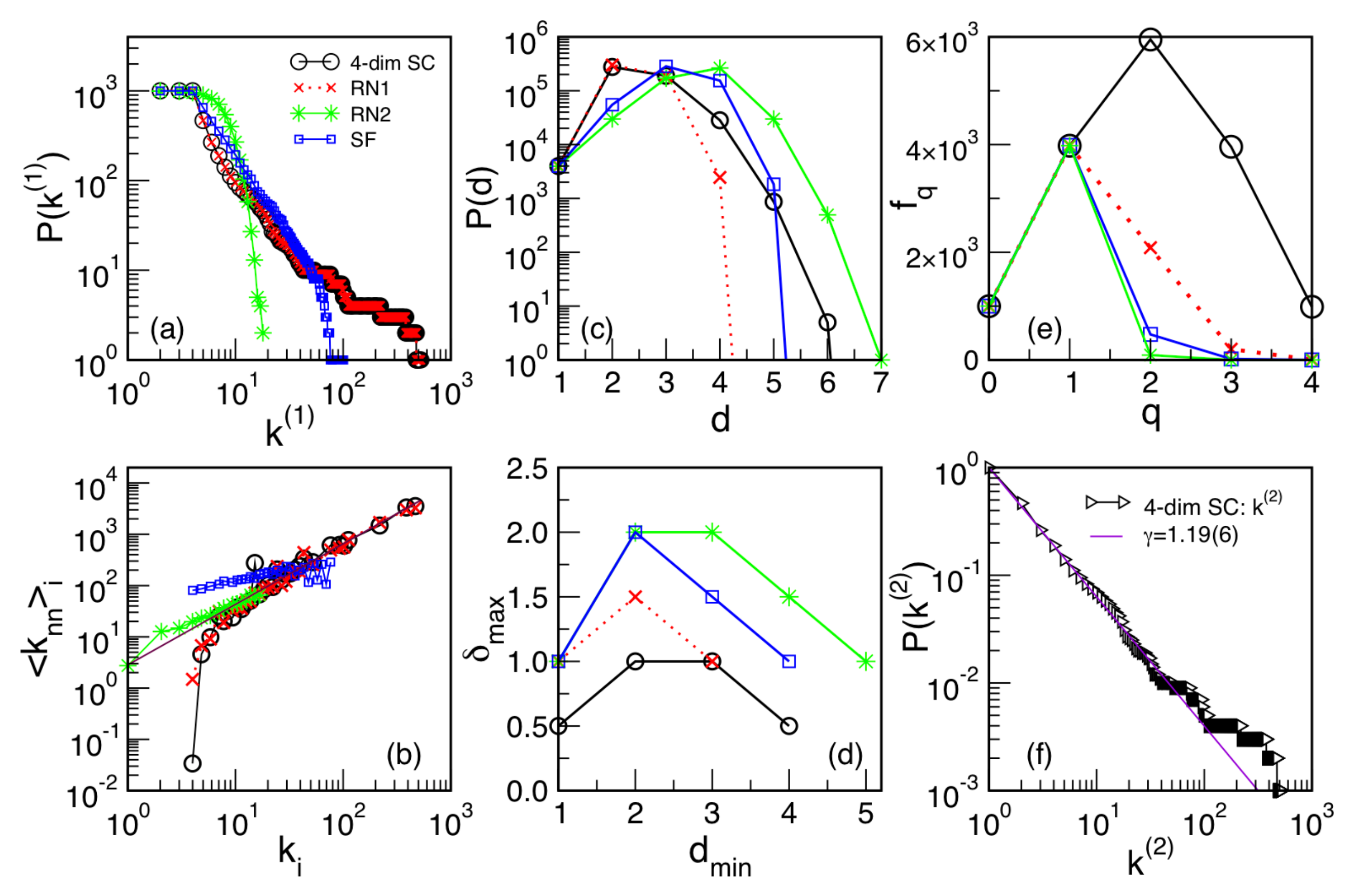}}\\
\end{tabular}
\caption{Visualisation of the homogeneous 4-dimensional simplicial complex SC of 1000 nodes (top, left) and the structure randomised to preserve the nodes degree, RN1 (top, right).  Panels a-e show different structural
  properties, in particular, the cumulative distribution of the node's
  degree (a), assortativity (b), the distribution of the shortest path
  distances (c) and the maximum hyperbolicity parameter (d), and the
  number of faces of different orders $q=0,1,2\cdots 4$ up to the
  maximal cliques (e). Different symbols (colours) are for the original
  4-dimensional SC, and the networks of the same size with the degree-preserving randomised structure RN1, the fully randomised structure RN2, and the simple scale-free network with the matching slope, SF. The same legend applies to panels (a--e). Panel (f) shows the normalised distribution of the number of triangles per node (generalised
 degree) in the 4-dimensional SC.   
}
\label{fig:nets}
\end{figure*}
\section{Network geometry underlying  synchronisation processes\label{sec:nets}}
As mentioned in the Introduction, we use the algorithm of
cooperative self-assembly introduced in ref.\ \cite{SC_we_SciRep2018,we-ClNets-applet} to
grow a simplicial complex (SC) by an assembly of 5-cliques. The geometric compatibility of its faces determines the attachment rules of a new
clique to the growing structure with the currently built-in cliques; besides, the chemical affinity parameter $\nu$ modulates the probability of
binding along a $q$-dimensional face. Effectively, see
\cite{SC_we_SciRep2018,SC_we_PRE2020}, the positive affinity parameter, as we used here $\nu=+5$, lead to the preference of sharing the largest sub-cliques. Thus, the newly added clique mainly consists of a new node and a  4-clique face shared with the previously added clique. An example of the resulting compact structure consisting of 5-cliques is shown in  Fig.\ \ref{fig:nets} top-left. To assess the relevance of a particular network property in the observed collective dynamics, we also study the synchronisation of phase oscillators on two randomised versions of our simplicial complex. Specifically, we perform random rewiring that preserves the degree of each node (the network is also shown in Fig.\ \ref{fig:nets}top-right) and a fully randomised structure. For further comparison, we also consider a simple scale-free network with the power-law exponent that coincides with the slope observed in the degree distribution of the SC for the intermediate degree, cf. Fig.\
\ref{fig:nets}.

Besides the high spectral dimension of our SC, $d_s\geq 4$ shown in \cite{SC_we-PREspectra2019}, several other structural measures of these networks relevant to the synchronisation dynamics are given in  Fig.\ \ref{fig:nets}a-f.  
Fig.\ \ref{fig:nets}e shows that, even though they all contain
the same number of nodes ($q$=0) and a similar number of edges ($q$=1 simplexes), they
significantly differ in the presence of higher simplexes.
Specifically, a small number of triangles ($q=2$ simplexes) as the
highest structures appear in the entirely random graph. Similarly, the
simple scale-free network possesses a small number of triangles and no
higher structures. On the other hand, the degree-preserving randomised
network still possesses about 30\% of the triangles compared to the
original simplicial complex and a few tetrahedrons ($q$=3
simplexes). Meanwhile, the number of tetrahedrons and 5-cliques in the
original complex is comparable to the number of edges and nodes,
respectively. The distribution of the number of triangles in which a
given node participates, also known as generalised degree $k_i^{(2)}$,
of our simplicial complex is shown in the panel (f) of Fig.\
\ref{fig:nets}. Besides the four hubs with many triangles
attached to them, the remaining part of the distribution obeys an algebraic decay with the increasing $k_i^{(2)}$.

At the level of edges, the underlying graph of our simplicial complex
exhibits some characteristic features  depicted in Fig.\
\ref{fig:nets}a-d in comparison with the other three
structures. Specifically, it exhibits a wide range of the degree
$k_i\equiv k_i^{(1)}$ with a few large-degree nodes. In the intermediate range,
the cumulative degree distribution has a power-law decay with the exponent
$\gamma \sim 1.81\pm 0.05$,  matched by the generated scale-free network, see Fig.\ \ref{fig:nets}a.
Naturally, the degree-preserved randomised structure obeys the same
degree distribution; meanwhile, the exponentially decaying
distribution characterises the fully randomised structure.
Moreover, our network possesses the assortative mixing among the
neighbouring node's degree \cite{sergey-lectures}; it is quantified by the
positive exponent $\mu >0$ in the expression $<k_{nn}>_i\sim k_i^\mu$
for the average degree of the neighbours of a node $i$ as a function
of the node's $i$ degree, suggesting that the nodes of similar
connectivity are mutually connected.  Fig.\ \ref{fig:nets}b shows the
assortative feature with $\mu\sim 1.19\pm 0.06$ for the graph of our simplicial complex. Notably, statistically similar assortative correlations are present in the degree-preserving randomised structure. Meanwhile, the random graph and the simple scale-free networks have $\mu \sim 0$ compatible with the absence of degree correlations.
Furthermore, Fig.\ \ref{fig:nets}c-d shows that, in the graph's metric
space (endowed with the shortest-path distance), these graphs have a
relatively small diameter, and hyperbolicity or negative curvature
\cite{Geometry_NegC_traffic,Hyperbolicity_TreeRG2012}, precisely 
they are $\delta$-hyperbolic with a small $\delta$ value
\cite{HB-BermudoGromov2013,HB-Bermudo2016}.  Moreover, due
to the attachments among cliques\cite{Hyperbolicity_cliqueDecomposition2017}, which are $0$-hyperbolic objects, it was shown \cite{SC_we_SciRep2018} that the topological graphs
of the emergent assembly are always $1$-hyperbolic.  Practically, this means that the maximum observed $\delta$ in the Gromov hyperbolicity criterion \cite{HB-BermudoGromov2013} can not exceed the value 1 for any four-tuple of nodes in that graph. In Fig.\ \ref{fig:nets}d we show how the $\delta_{max}$ can vary with the minimal distance in a large number of sampled four-tuples for all four network structures. Notably, $\delta_{max}=1$ for the graph of our
simplicial complex, as expected, and it increases by 1/2  with the
degree-preserving randomisation of edges.  In the small-$\delta$
graphs, such increases of the hyperbolicity parameter are attributed \cite{HB-Bermudo2016}
to the appearance of a characteristic subjacent structure, usually a
new cycle compatible with the new $\delta$ value. Our graph's complete
randomisation and the simple scale-free structure appear to possess
even larger cycles, resulting in the $\delta_{max}=2$. 
In the following two sections, we will investigate the synchronisation
processes among phase oscillators interacting via edges and triangles of these networks.

\section{Phase synchronisation with the competing simplex-based
  interactions\label{sec:dyn}}
The phase variable is an angle in 2-dimensional space,  $\theta_i$, associated with the
network's nodes $i=1,2,3,\cdots N$.  The local interactions among these dynamical variables of the strengths $K_q$ are provided by the network's
topology elements, which are strictly related to the corresponding
faces of the simplicial complex. In this work, we consider two leading
interactions associated with the edges ($q=1$) and
interactions among triplets located on a triangle face ($q=2$), with the
strength $K_1$ and $K_2$, respectively. The  evolution equations given by 
 \begin{eqnarray}
 \dot{\theta_i}  =  \omega_i + \frac{K_1}{k^{(1)}_i} \sum_{j=1}^{N}A_{ij} \sin{(\theta_j - \theta_i)} +  \nonumber \\
 +  \frac{K_2}{2 k^{(2)}_i} \sum_{j=1}^{N} \sum_{l=1}^{N} B_{ijl}\sin{(\theta_j + \theta_l - 2\theta_i)} \label{eq:dyn}
 \end{eqnarray}
are coupled via the two interaction terms. In particular, $A_{ij}$ is an element of the 1-simplex adjacency matrix $\mathbf{A}$, such that
$A_{ij}$ = 1 if nodes $i,j$ are conneted by a link, and 0 otherwise; meanwhile,
$ B_{ijl}$ is an element of the 2-simplex adjacency tensor $\mathbf{B}$,
such that $B_{ijl}$ = 1 if nodes $i, j, l$  belong to a common 2-simplex, and 0 otherwise.
The normalisation factors in eq.\ (\ref{eq:dyn}) are the respective simplex
degree of a node, $k_i^{(q)}$, i.e. number of distinct $q$-simplices
that node $i$ is part of. Specifically, $k_i^{(1)}$ is the number of
1-simplices (edges) incident on node $i$, and $k_i^{(2)}$ is the
number of 2-simplices (triangles) incident on node $i$.
Thus, equal weightage is given to all the terms contributing to the sum in each interaction term. It is important to note the number of edges and
the number of triangles per node   in the underlying graph of our simplicial
complex obey a broad (partly a power-law) distribution, cf.\ Fig.\
\ref{fig:nets}a,f.  
In eq.\ (\ref{eq:dyn}), $\omega_i$ is the intrinsic frequency of the $i$-th oscillator, which dictates its motion when there is no interaction with other oscillators in the network. The pairwise interactions seek to reduce the difference between the phase of the $i$-th oscillator and each of its neighbouring oscillators when $K_1>0$.  In contrast, the
oscillators tend towards opposite phases when $K_1<0$.
The third term, 
 representing three-node interactions of the $i$-th oscillator based on
 each 2-simplex incident on node $i$, is a natural
 generalization of the pair-wise interaction term \cite{DdimSC_boccaletti2020}.
It should be stressed that the interactions between these three nodes
occur over faces of the simplicial complex and not over any given
three nodes. Furthermore, this interaction term is symmetric in $i$,
in that it is unaffected by permutations in the other two
indices. Revealing the  impact of the 2-simplex term in Eq.\ (\ref{eq:dyn})
on the synchronisation processes that are promoted by the pairwise interactions is
 one of the objectives of this work.

As it is widely accepted, the degree of synchronisation of the whole
network is quantified by the Kuramoto order parameter 
\begin{equation}
r= \Big \langle \left|\frac{1}{N} \sum_{j=1}^{N} e^{i
    \theta_{j}}\right| \Big \rangle \ ,
\label{eq:op}
\end{equation}
where the brackets $\langle \cdot \rangle$ indicate  the time average.
Hence, $r=1$  represents the perfect synchronisation, i.e., all phases
are equal, and $r=0$ in the disordered phase.  Meanwhile, the stable states with 
$0<r<1$ indicate the presence of more complex  patterns and partial synchronisation.
 
 In the simulations, the initial conditions are set for $\theta_i$ and
$\omega_i$ as the uniform random number in the range
$\theta_i\in [0,2\pi]$ and the Gaussian random number with a zero
mean and unit variance, respectively.  
The numerical solution of the set of equations
(\ref{eq:dyn}) is performed using a numerical integrating function
\texttt{odeint} from Python's SciPy library
\cite{2020SciPy-NMeth}. This function integrates a system of ordinary
differential equations (ODEs) using the \texttt{lsoda} solver from the
Fortran library ODEPACK. It solves ODEs with the Adams
(predictor-corrector) method the Backward Differentiation Formula for
non-stiff and stiff case, respectively. For each set of parameter
values, the system is iterated for $50 000$ steps. The last $20 000$
iterations are used to calculate the order parameter as in Eq.\
(\ref{eq:op}). Further, in order to study hysteresis, we track the
trajectory of the system as the coupling parameter $K_1$ is first
adiabatically increased and then decreased. The step size of coupling
$K_1$ is taken to be $0.1$.  Alternatively, $K_2$ is varied in a suitable range,
meanwhile fixing  several representative  $K_1$ values, as described
in the second part of 
sec.\ \ref{sec:HL}. A detailed program flow is given in the Appendix.

\subsection{The case $K_2=0$: Synchronisation under exclusively pairwise
  interactions\label{sec:k2zero}}
Before considering the competing simplex-based interactions, we will describe the synchronisation process under the
pairwise interactions alone, i.e., when $K_2=0$ in Eq.\ (\ref{eq:dyn}). As mentioned above, these interactions
are enabled by the edges of the substrate network, which is the
1-skeleton of our 4-dimensional simplicial complex. Hence, different network features from local to global level
are expected to play a role in the cooperative
behaviours, depending on the interaction strength $K_1$. 
\begin{figure}[!htb]
\begin{tabular}{cc} 
\resizebox{20pc}{!}{\includegraphics{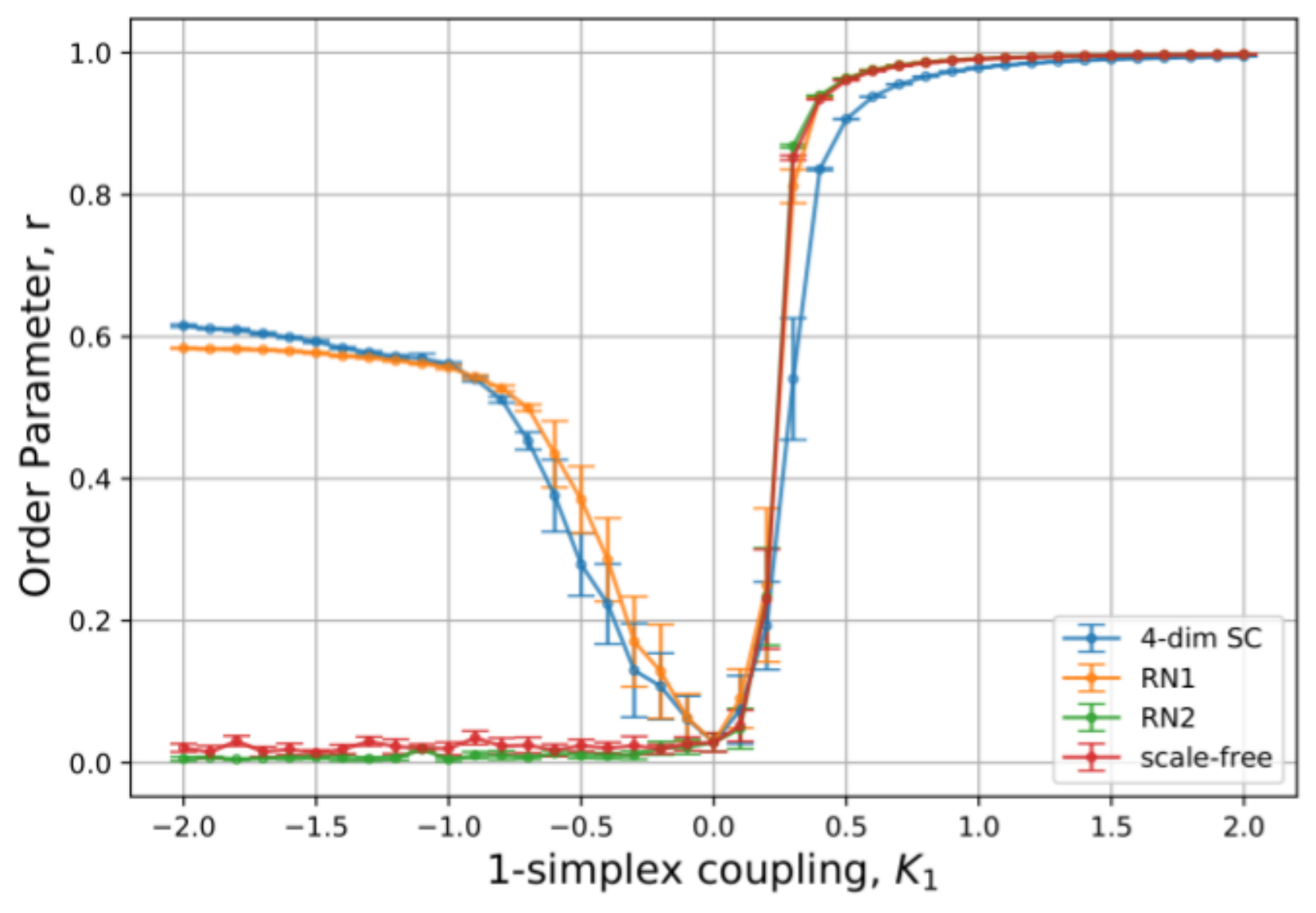}}\\
\end{tabular}
\caption{The order parameter $r$ as a function of the 1-simplex
  coupling strength $K_1$ without the  higher-order
  interactions. Calculations are plotted for 4-dimensional simplicial
  complex (blue), randomized network with the original degree
  distribution (orange), fully randomized network (green), and
  a simple scale-free network with the same number of nodes and edges (red). }
\label{fig:opK2zero}
\end{figure}
For the $4$-dimensional simplicial complex, as $K_1$ is increased from
zero up to $K_1 = 2.0$, and we observe a continuous  transition from a
desynchronized state ($r \approx 0$) to a completely synchronized state
($r \approx 1$), see Fig.\ \ref{fig:opK2zero}. On the other hand, with
the
negative values of $K_1$ decreased from $K_1=0$  to $K_1 = -2.0$,  the
order-parameter increases in a different manner and reaches the value
$r \approx 0.6$, comprising a partially synchronised state. To demonstrate what
network property can be responsible for the observed synchronisation
properties, we performed the simulations on the two randomised
versions of the network, as described in Section\ \ref{sec:nets}.
Notably, for the degree-preserving randomised structure, RN1,
qualitatively similar behaviour of the order parameter is found. In contrast, for the fully randomised network, the order parameter remains zero for all values of $K_1\leq 0$. Interestingly, almost identical values of the order-parameter compatible with the absence of synchronisation at
negative pairwise interactions are found
in a simple scale-free network, as
shown in Fig.\ \ref{fig:opK2zero}. Therefore, we can conclude that the
node's assortative degree correlations  in the network of our simplicial complex
and in the corresponding degree-preserving randomised version, cf.\ Fig.\ref{fig:nets}b, can be
responsible for the appearance of the partial synchronisation for the
negative pairwise interaction. 

\begin{figure*}[!htb]
\begin{tabular}{cc} 
\resizebox{34pc}{!}{\includegraphics{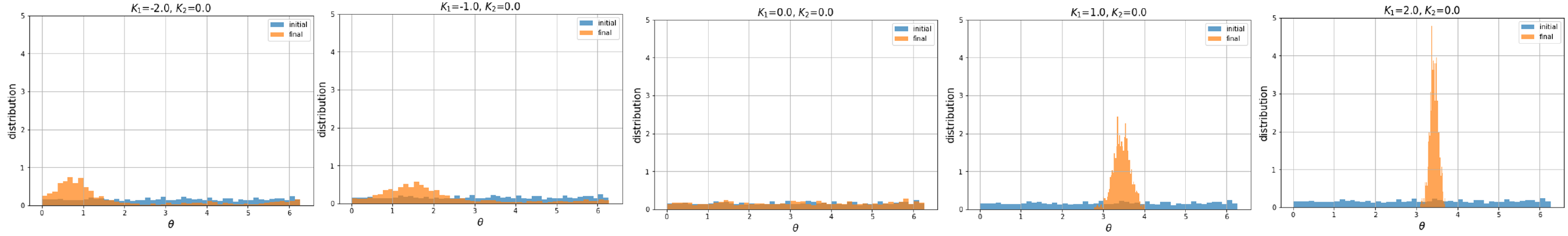}}\\
\resizebox{34pc}{!}{\includegraphics{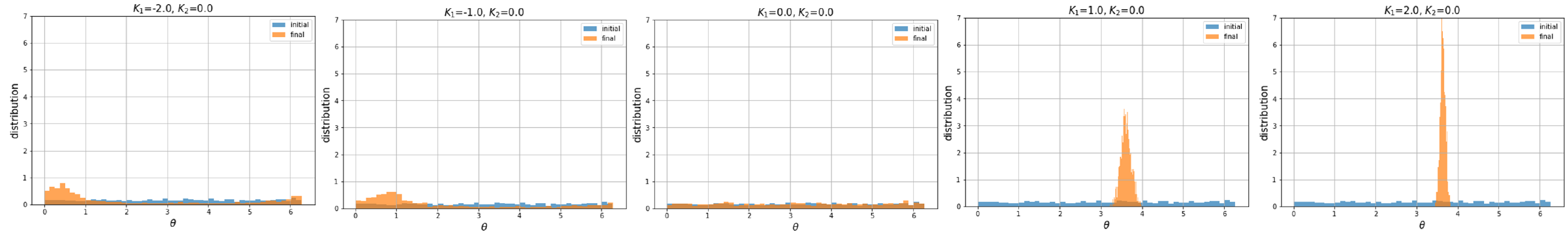}}\\
\resizebox{34pc}{!}{\includegraphics{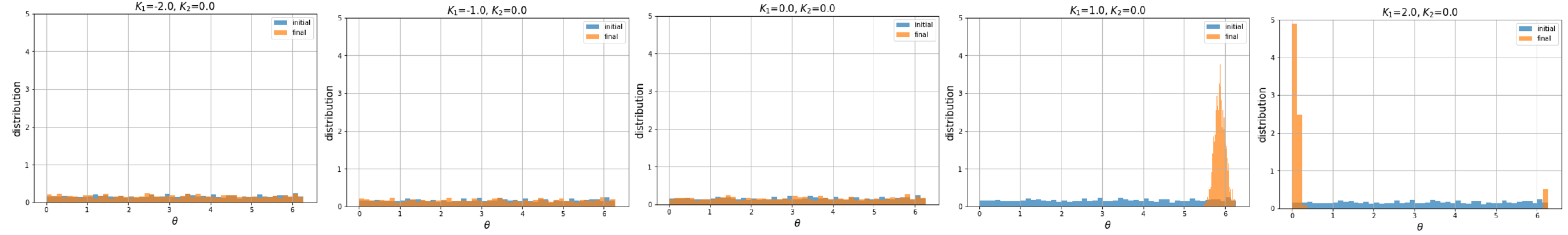}}\\
\resizebox{34pc}{!}{\includegraphics{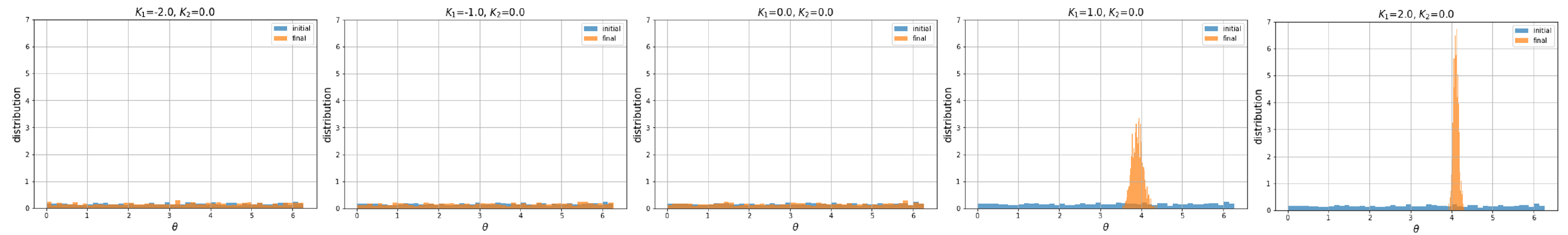}}\\
\end{tabular}
\caption{Distribution of phases (measured in radians) in the initial (blue) and final (orange)
  states of the synchronization simulations for different values of
  the pairwise interaction for four network structures described in
  sec.\ \ref{sec:nets}. The first raw of panels is for the network of the
  4-dimensional simplicial complex, the second raw is for the
  randomized network with the original degree distribution (RN1), the
  third raw is for the fully randomized network (RN2), and the last
  raw is for the scale-free network with the same number of nodes and edges (SF).}
\label{fig:phases}
\end{figure*}

The distribution of phases over nodes, or the synchronisation pattern, is
expected to depend on the structure. Here, the histograms of phases corresponding to different representative values of the coupling
strength $K_1$ are shown in Fig. \ref{fig:phases} for different network
structures. While the order parameter in the scale-free and entirely random network is practically identical, cf.\ Fig.\ \ref{fig:opK2zero}, the phases in the synchronised state appear to be different, which can be attributed to the degree distribution as the only measurable difference between these networks. On the other hand, there is a remarkable similarity in the distribution of phases in the simplicial-complex network and its degree-preserving randomised version. Moreover, the
peak for large positive values of  $K_1$ is close to the one seen
in the corresponding scale-free structure. On the negative $K_1$ side, the
majority of phases also appear to be in the same region; see the top two raws of
Fig.\ \ref{fig:phases}. How the partially
synchronised state in these correlated networks appears is another
important issue. We anticipate that a large number of triangles, as
shown in Fig.\ \ref{fig:nets}e, can be responsible for the frustration
effects leading to the partially synchronised states in these two
networks.
In the following, we will examine the impact of the triangle-based
interactions in our simplicial complex.

\begin{figure*}[!]
\begin{tabular}{cc} 
\resizebox{34pc}{!}{\includegraphics{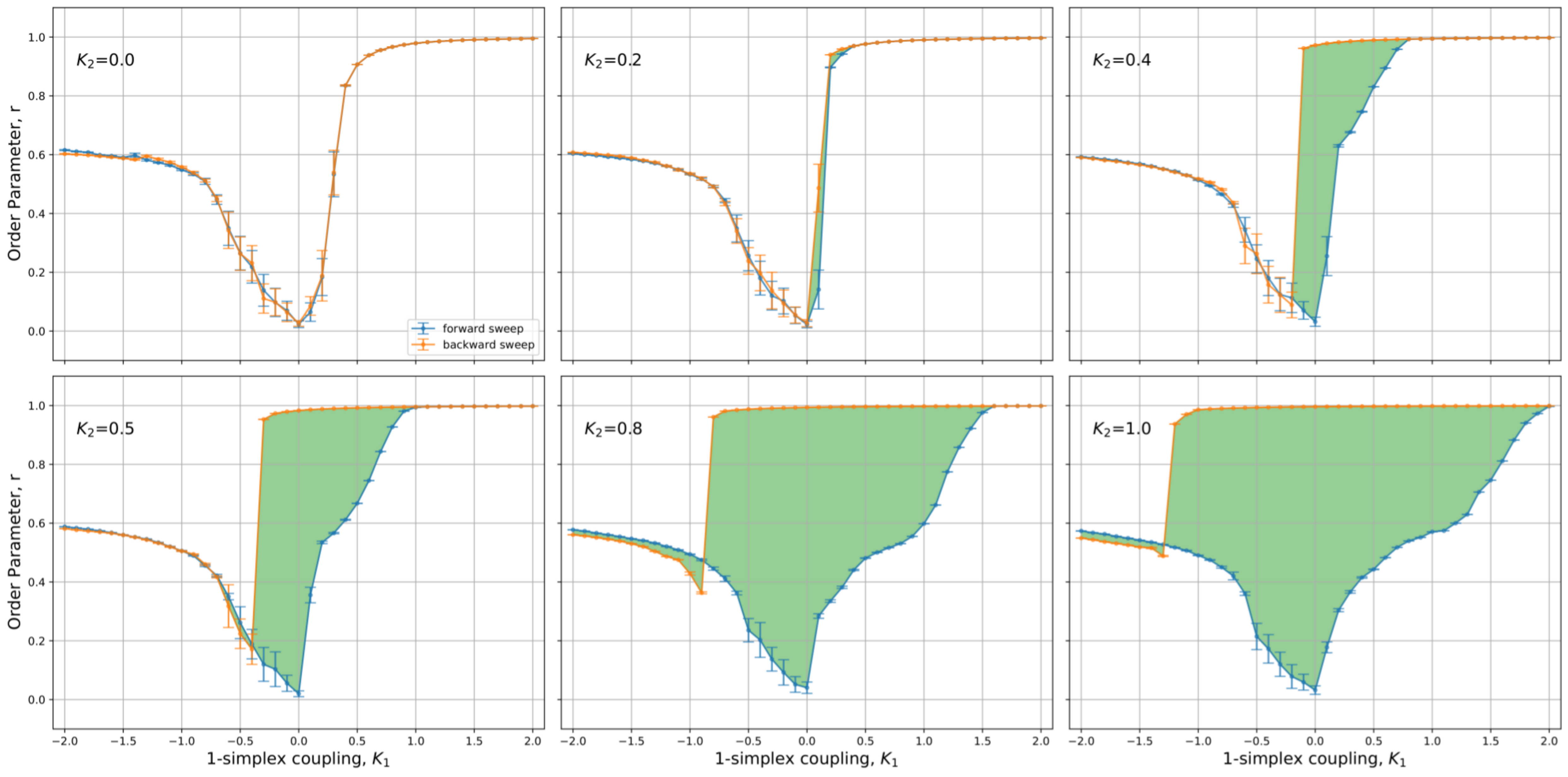}}\\
\end{tabular}
\caption{Synchronization with higher-order interactions: Hysteresis sweep of the order parameter as a function of 1-simplex coupling strength $K_1$ for different 2-simplex coupling strength $K_2$ values. As the value of $K_2$ increases, we notice an increase in the size of the hysteresis loop.}
\label{fig:HLs}
\end{figure*}

\section{Hysteresis loop induced by higher-order
  interactions\label{sec:HL}}
\begin{figure}[!htb]
\begin{tabular}{cc} 
\resizebox{20pc}{!}{\includegraphics{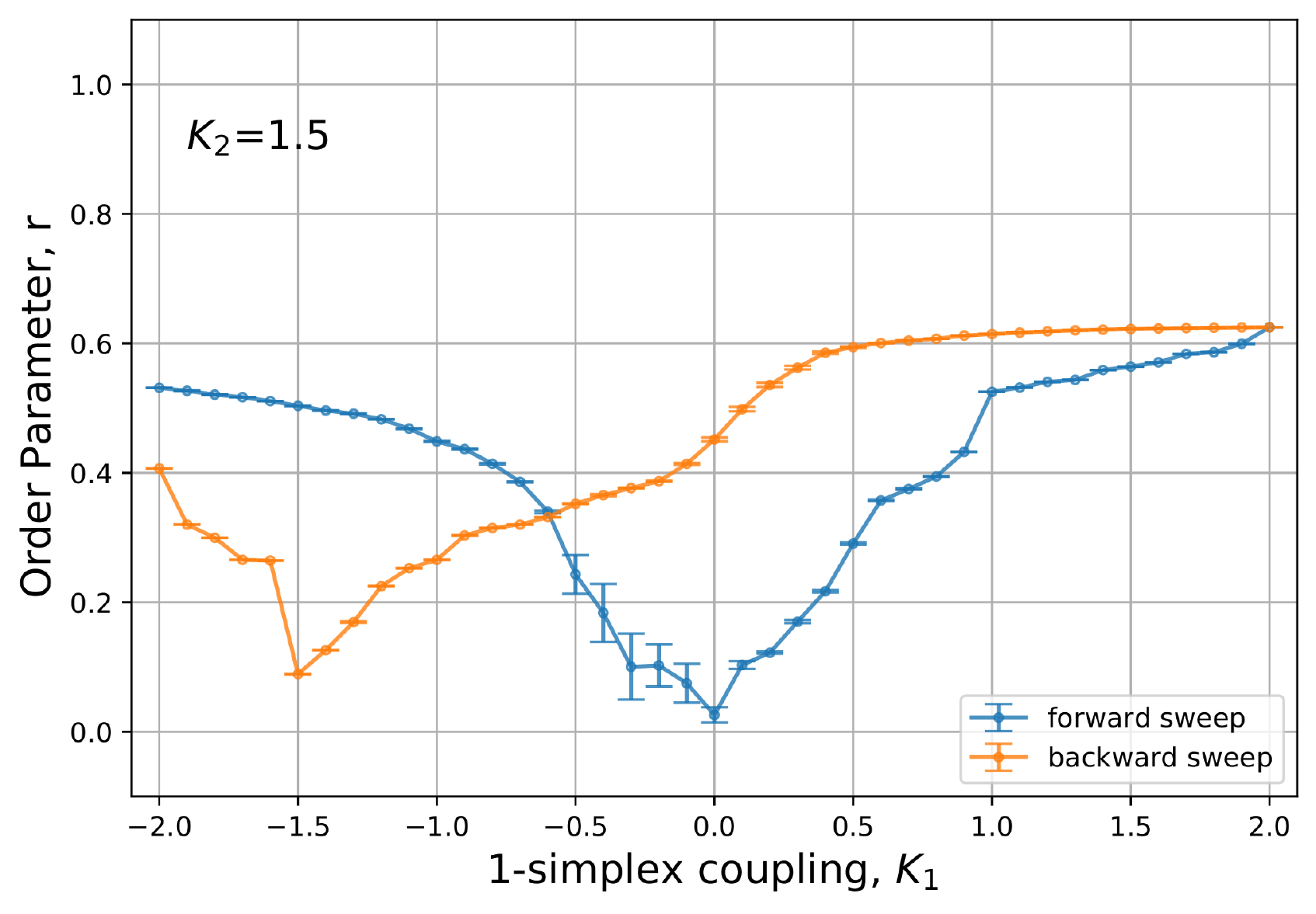}}\\
\end{tabular}
\caption{Hysteresis loop for strong 2-simplex interaction $K_2$:  only partial synchronisation is accessible even at a large positive $K_1$; the abrupt desynchronisation disappears.}
\label{fig:HLlargeK2}
\end{figure}

\begin{figure}[!htb]
\begin{tabular}{cc} 
\resizebox{18pc}{!}{\includegraphics{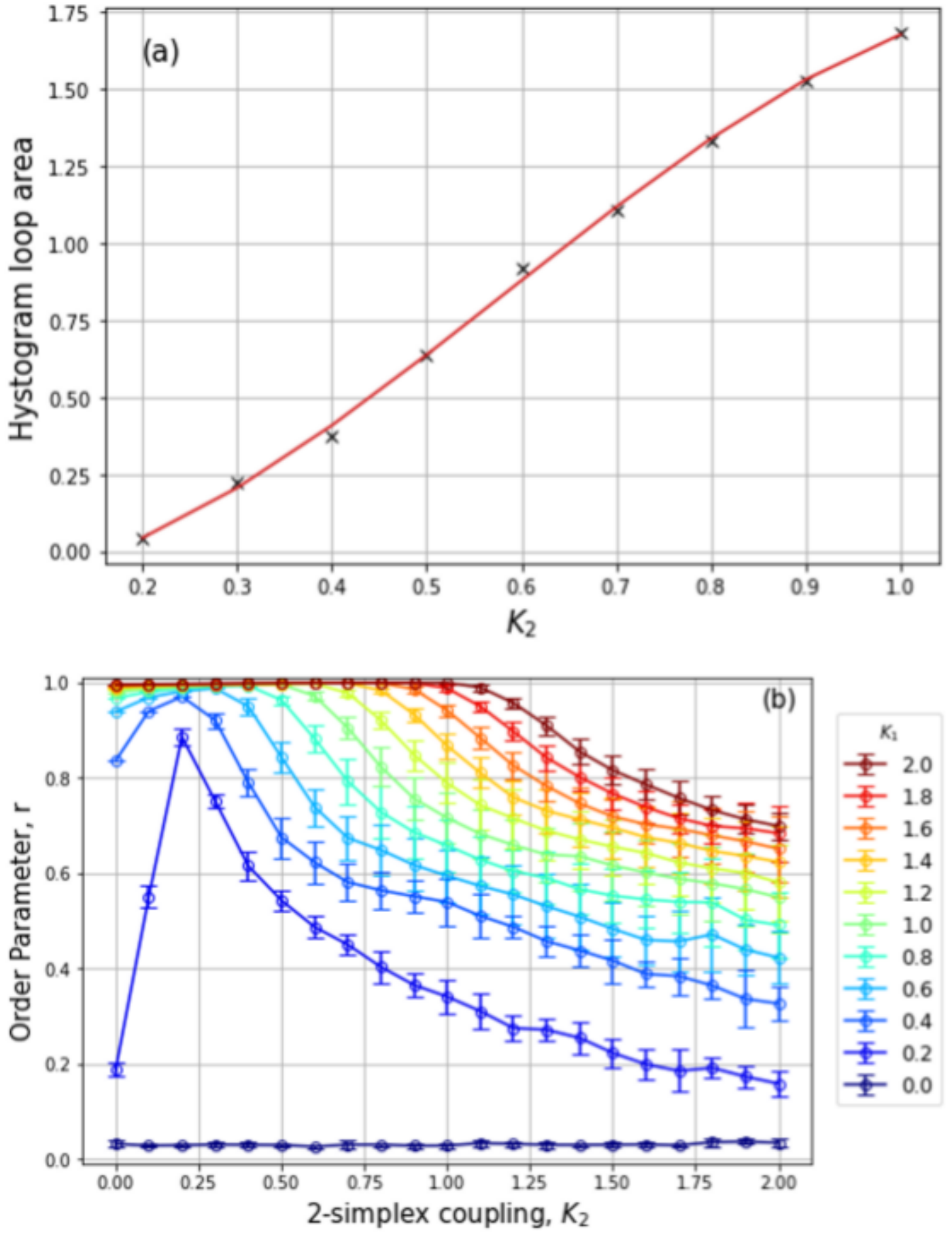}}\\
\end{tabular}
\caption{  (a) The size of the hysteresis loop obtained by
  increasing and then decreasing pairwise
  interaction $K_1$, as shown in Fig.\ \ref{fig:HLs},  plotted against $K_2$; the parameters of the cubic polynomial fit are
  given in the text. (b) The order
  parameter $r$ against $K_2$ for different
  values of $K_1$ indicated in the legend.}
\label{fig:_K2}
\end{figure}

In this section, we will focus attention on synchronization dynamics
on our network in the presence of triangle-based
interactions,  i.e., using $K_2\neq 0$ in
equation (\ref{eq:dyn}). We plot the time averaged
order parameter $r$ as a function of $1$-simplex coupling $K_1$, for
different $2$-simplex coupling values $K_2 = 0.0, 0.2,  0.4, 0.5,
 0.8, 1.0$ in Fig. \ref{fig:HLs}. 
 In these plots, first $K_1$ is increased adiabatically from $K_1 = $
 -2.0 to +2.0 (forward sweep), and then decreased from $K_1 =$ +2.0 to -2.0 (backward sweep) in steps
of $dK_1=$0.1. For each plot shown in Fig. \ref{fig:HLs}, we see that at $K_1=$-2,
the system is partially synchronised, with a finite $r$ value. The
system gradually desynchronises as $K_1$ tends towards zero. Next, as
$K_1$ is increased towards $+2.0$, a continuous increase of the
level of synchronisation occurs. For higher values of $K_2$, the
transition remains continuous but slows down due to possible
phase-locked configurations in the region around $K_1\sim 0.5$, 
 and gradually reaches the complete synchrony at larger values of $K_1$.
In the backward sweep, as $K_1$ decreases from $+2.0$ to zero and
further towards $K_1=-2$, the desynchronisation transition largely
depends on the value of the triangle-based interactions. Namely, 
when $K_2 = 0$, the  transition is continuous  following the same trajectory as the forward transition,  through the fully
desynchronised state at $K_1=0$, and ending up with the partially
synchronised state at $K_1=-2$. 
As the coupling $K_2$ is increased, the forward and
backward transitions are no longer reversible. 
More precisely, when the 2-simplex
interaction strengths exceeds  $K_2 \gtrsim $ 0.4, the level of synchronisation
slows down in the forward sweep due to  phase-locked configurations,
as mentioned above.
Meanwhile, in the backward sweep, we note a
discontinuous desynchronisation decay towards the $K_1 < 0$
branch.
The occurrence of an abrupt desynchronisation has been
previously reported in
\cite{Sync_Arenas_PRL,DdimSC_boccaletti2020,SC_Arenas_NatureComm} as a prominent effect of higher-order interactions with different coupling
types. In this context, the abrupt destruction of the synchronised state in our simplicial complex is also expected. 
What is new is the specific dependence of the hysteresis loop and thus
the abrupt desynchronisation phenomenon on the strength of the
2-simplex interactions, as demonstrated by the results in
Fig.\ \ref{fig:HLs} and Fig.\ \ref{fig:HLlargeK2}. The abrupt
transition is to a partially synchronised state with a non-zero value
of the order parameter. Notably, an abrupt
decay of the completely synchronised state occurs when $K_2$ is large
enough to balance the effects of the non-positive pairwise
interaction  $K_1\leq 0$. Thus, the only complete desynchronisation transition appears at the point $K_1=0$ for a small $K_2$ value, as Fig.\
\ref{fig:HLs} shows. 
Fig. \ref{fig:HLs} shows that, beyond this value of $K_2$,
the hysteresis loop grows in size as $K_2$ is increased, affecting
both the forward sweep at the positive $K_1$ side and the size of the
first-order jump.  This scenario continues for a wider range of values
of the 2-simplex interaction strength as long as the large positive
pairwise interactions are sufficient to maintain a complete
synchrony. However, for larger values of $K_2$, the fully synchronised state is no longer accessible; instead, a kind of partially
synchronisation is reached under the competing interactions. The
backward sweep from such a state, as shown in Fig.\
\ref{fig:HLlargeK2}, closes up an entirely different shape of the
hysteresis loop appears with
two distinct parts at positive and negative $K_1$, and continuous
changes of the order parameter.  Hence, we can conclude that the
impact of the 2-simplex encoded interactions on our 4-dimensional
simplicial complex strongly depends on the sign and strength of the
pairwise interactions. An overview of its effects is
displayed in Fig.\ \ref{fig:_K2}, and discussed in sec.\ \ref{sec:concl}.

Furthermore, we analyse how the hysteresis loop area grows with
increasing $K_2$. Particularly, we plot the area of the hysteresis
loops for different values of $K_2$ against $K_2$ in
Fig. \ref{fig:_K2}a.  The curve is best fitted with a cubic function $f(x)=-2.3767 K_{2}^3 + 4.1631 K_{2}^2 - 0.0050 K_{2} - 0.1043$ with root-mean-square error $0.01866$.

We highlight the competing nature of the $1$-simplex and
$2$-simplex interactions. To that end, we carry out synchronisation
simulations for different pairs of coupling strengths $K_1$ and
$K_2$, illustrated  by plotting the order parameter, $r$, as a
function of $2$-simplex coupling $K_2$ in figure \ref{fig:_K2}b. 
We notice that the order parameter remains negligible if the pairwise
interactions are absent,  $K_1=0$,  for all values of $K_2$
considered, suggesting that the 2-simplex interactions alone can not
induce the system's synchronisation.
Next, we notice that for finite but low values of $K_1$, increasing
$K_2$ leads to higher synchronisation, but only until around $K_2 \leq
0.2$. A further increase in $K_2$ leads to decreasing the system's
level of synchronisation, as seen from the decaying values of
$r$ with $K_2$. Note that higher $K_1$ are gradually needed
to compete with the desynchronising effects due to high $K_2$.  Subsequently, the curves of $r$ in figure \ref{fig:_K2}b decrease with the increasing $K_2$.  In this range of  $K_2$ values, the complete synchrony is no longer accessible, as also
demonstrated with the hysteresis loop in Fig.\ \ref{fig:HLlargeK2}.

\section{Discussion and Conclusions\label{sec:concl}}
How high-dimensional simplicial complexes can shape
the dynamics is a question of great relevance to many functional systems. Among these, a prominent example is the human connectome structure underlying the brain functional complexity. 
To address this issue, we have studied the processes of
phase synchronisation  on a homogeneous 4-dimensional simplicial complex of a given size ($10^3$ nodes); we have considered the leading
interactions encoded by 1-simplex (edges) and 2-simplex (triangles)
faces and varying the respective strengths $K_1$ and $K_2$. Our
results revealed a variety of scenarios for the synchronisation and desynchronisation (both to complete and partially desynchronised states), depending on the sign of the pairwise interactions and the
geometric frustration promoted by the triangle-based interactions.
The latter can be attributed to the actual organisation of 5-cliques that
make the simplicial complex; notably,  every link in this complex is
a shared face of at least three triangles. Moreover, the geometrical properties of its underlying graph play their role. These are the assortative degree correlations, the graph's hyperbolicity, and the high spectral dimension.
 
More precizely, we have demonstrated that:
\begin{itemize}
\item the 1-simplex interactions of both signs $K_1\gtrless 0$ promote the
  synchronisation but with different mechanisms; no synchrony can
  arise due to 2-simplex interactions alone;

\item for $K_1>0$ the two interaction types have competing effects,
  and the complete synchrony can be reached for a moderate range of $K_2$, balanced by the increasingly stronger pairwise coupling $K_1$; 

\item for the negative pairwise interactions $K_1<0$, however, the
  2-simplex interactions support the mechanisms leading to partially
  ordered states due to $K_1$; 

\item the prominent impact of the 2-simplex interactions is seen in
  the opening-up of the hysteresis loop and the appearance of a finite
  jump  in the backward sweep starting from the completely
  synchronised state, in analogy to the abrupt
  desynchronisation found in other studies
  \cite{Sync_Arenas_PRL,SC_Sfboccaletti2021,SC_Arenas_NatureComm}. Note
  that, in our case, the desynchronisation is partial following the
  forward branch for $K_1\leq 0$, except when it occurs precisely at the point
  $K_1=0$;  

\item eventually, for substantial 2-simplex interactions $K_2 \gtrsim K^*$ the geometric frustration  prevailed, leading to partial
  synchronisation even though a large positive $K_1$ is adequate; the
  abrupt desynchronisation entirely disappears; on
  the negative $K_1<0$ side, a new segment of the hysteresis loop
  opens up, suggesting that potentially different orderings in the
  frustrated synchronisation may be competing before a significant negative $K_1$ prevails; 
\end{itemize}

Our 4-dimensional simplicial complex with the
simplex-encoded interactions represent an excellent example to investigate how geometry influences the synchronisation and desynchronisation processes on it. 
Even though the considered higher-order interactions are the leading
cause of new dynamical phenomena, the collective
behaviour's genesis is rooted in the pairwise interactions. Hence, certain
nontrivial features of the underlying network are highly relevant.
Our study sheds new light on the competing role of simplex-embedded
interactions in high-dimensional simplicial complexes, which occur in
many natural dynamical systems. In this context, some outstanding
questions remain for the future study.  For
example, one such question regards the relative importance of the order of interactions that can be embedded in a given simplicial complex. Moreover, our approach traces the ways to study the role of defect simplexes, the temporally-varying simplicial architecture and distributed weights, which can have profound effects on collective dynamic behaviours.

\section*{Acknowledgments}

B.T.  acknowledge the financial support from the Slovenian
Research Agency under the P1-0044 program. 
N.G. thanks IIT Madras for the CoE Project
No. SP20210777DRMHRDDIRIIT. 
M.C. acknowledges the use of the computing resources at HPCE, IIT Madras.


\section*{Appendix}
\
\subsection{Program Flow\label{alg}}
\begin{algorithm}[H]
\label{alg:hl}
\begin{algorithmic}[1]
\STATE {INPUT Graph $\mathcal{G}$, stored as an list of edges and list
  of triangles;}\\

\STATE Initialize phases of the oscillators, $\theta_i$, such that they are distributed uniformly between $0$ and $2 \pi$;\\

\STATE Initialize intrinsic frequencies of the oscillators, $\omega_i$, such that they are distributed normally, with zero mean and unit variance;\\

\STATE Set the value $K_2$. Assign $K_2 = 0$, if $2$-simplex interactions are to be ignored.
Assign finite $K_2$, if $2$-simplex interactions are to be added;\\

\STATE Set the value $K_1 = -2.0$; Set the incremental change in $K_1$ to be $dK_1 = 0.1$;\\

\STATE{\tt{Forward sweep:}}

\WHILE{$K_1 \leq  +2.0$}
{
    \FORALL{nodes $i \in \mathcal{G}$}
    {
        \STATE solve the differential equation (1);
    }
    \ENDFOR
    \STATE Calculate order parameter for the system, using equation (3);
    
    \STATE Increase $K_1$ by $dK_1$;
}
\ENDWHILE
\STATE{\tt{Backward sweep:}}

\WHILE{$K_1 \geq  -2.0$}
{
    \FORALL{ nodes $i \in \mathcal{G}$}
    {
        \STATE solve the differential equation (1);
    }
    \ENDFOR
    \STATE Calculate order parameter for the system, using equation (3);
    
   \STATE  Decrease $K_1$ by $dK_1$;
}
\ENDWHILE
\STATE{END}
\end{algorithmic}
\end{algorithm}


\end{document}